\documentclass{pazh}
\usepackage{graphicx}

\newcommand{\km}{\,\mbox{km}\,\mbox{s}^{-1}}
\def\Ha{\hbox{H$_\alpha$\,}}
\def\Hb{\hbox{H$_\beta$\,}}
\def\farcm{\hbox{$.\mkern-4mu^\prime$}}
\def\farcs{\hbox{$.\!\!^{\prime\prime}$}}

\begin{document}

\title{A Spectroscopic Study of the Peculiar Galaxy UGC~5600
}

\author{
L.V. Shalyapina\inst{1}\and A.V. Moiseev \inst{2}\and V.A.Yakovleva\inst{1}
}

\institute{
Astronomical Institute, St. Petersburg State University, 
Bibliotechnaya pl.2, Petrodvorets, 198904 Russia
\and
Special Astrophysical Observatory RAS, 
Nizhni\u{\i} Arkhyz, Karachaevo--Cherkesia, 357147 Russia
}

\offprints{L.V.~Shalyapina, \email{lshal@astro.spbu.ru}}
\date{Received Feb 8, 2002 }

\titlerunning{A Spectroscopic Study of the Peculiar Galaxy UGC~5600}
\authorrunning{ L.V.~Shalyapina\inst{1} et al.}

\abstract{
We present our observations of the galaxy UGS 5600 with a long-slit spectrograph (UAGS) 
and a multi-pupil field spectrograph (MPFS) using  the 6-m telescope of  Special Astrophysical Observatory.
Radial-velocity fields of the stellar and gaseous components were constructed for the central
region and inner ring of the galaxy. We proved the existence of two almost orthogonal kinematic subsystems
and conclude that UGC 5600 is a galaxy with an inner polar ring. In the circumnuclear region, we detected
noncircular stellar motions and suspected the existence of a minibar. The emission lines are shown to
originate in H II regions. We estimated the metallicity from the intensity ratio of the [NII]$\lambda6583$ and \Ha 
lines to be nearly solar, which rules out the possibility that the polar ring was produced by the accretion of
gas from a dwarf companion. c 2002 MAIK "Nauka/Interperiodica ". 
}

\maketitle

\section{INTRODUCTION}

Polar-ring galaxies (PRGs)  are objects with two kinematic subsystems rotating in roughly orthogonal
planes.  Based on its structural features (see Fig. 1), Whitmore et al. (1990) listed the peculiar galaxy
UGC 5600 among the most probable PRG candidates:  its amorphous main body is surrounded by a
broad outer ring and the brightenings attributed to the inner ring are observed in the EW direction
on both sides of the center at a distance of $\sim10''$.  
UGC 5600 is a member of a double system (VV 330).   Its companion, the galaxy UGC 5609, is at 1\farcm4 
(about 15 kpc in projection onto the plane of the sky)  to the
southeast and has a similar radial velocity.  Recently 
two other galaxies with similar redshifts have been
detected near VV 330 (Galletta et al. 1997) ;  all of them may represent a group of galaxies.  

Among other galaxies from the catalog of Whitmore et al. (1990)  , UGC 5600 was observed in the radio range at a wavelength of 21 cm (Richter et al.  1994).  The $M_{HI} /L{B}$ ratio was found to be 0.86 , characteristic of late-type galaxies. CO-line observations revealed molecular hydrogen in all structures of the
galaxy (Galletta et al. 1997).   

The spectra of the galaxy along its major and minor axes were obtained by Reshetnikov and Combes
(1994).  The radial-velocity curves are complex in shape. The authors suggested counterrotation in the central part of the galaxy ($ r<5''$ )  and gas rotation around its major axis. In addition, they pointed out that the galaxy is rich in gas and that the \Ha  emission
extends to $30''$ (6 kpc). 

 A detailed photometric study of UGC 5600 (Karataeva et al., 2001) shows that this is most likely a
late-type spiral (Scd)  galaxy with an inner polar ring, which is projected onto the galaxy main body on the
northern side and is seen through it on the southern side. The structure that was taken in the catalog of
Whitmore et al. (1990)  as an outer ring represents two tightly wound spiral arms. 

The final conclusion whether UGC 5600 belongs to PRGs can be reached only after proving
the existence of two nearly orthogonal kinematic subsystems. 

The distance to the galaxy is 37.6 Mpc ($H_0 = 75 \km \mbox{Mpc}~{-1} $)  and the scale is 0.18 kpc in $1''$ . 

\section{OBSERVATIONS AND DATA REDUCTION}

All of the spectroscopic data were obtained at the 
prime focus $(F/4 ) $ of the 6m telescope of Special Astrophysical Observatory (SAO). A log of observations
is given in the table~1. 

The observations with the UAGS long-slit spectrograph (Afanasiev et al., 1995)  were carried out
in January 2000 at two slit positions: along the galaxy major axis ($PA =0^\circ  -2^\circ$)  and along the polar-ring 
major axis ($PA =85^\circ$) ; according to Karataeva et al. (2001),  this axis passes $2''$ south of the galaxy 
photometric center (see Fig. 1).  The spectral range covered included the \Ha, [N II ]$\lambda6548,6583$ , and
[S II ]$\lambda6716, 6730$ emission lines. The detector was a Photometrics $1024\times1024$ -pixel CCD array. In the
observations, the spectrograph slit size was $2''\times140''$ , the reciprocal dispersion was 1.2\AA per pixel, 
the spectral resolution was 3.6\AA, and the angular scale along the slit was 0\farcs4 per pixel.

The UAGS spectra were reduced by using standard procedures from the ESO-MIDAS package. After the primary reduction, we carried out a smoothing along the slit with a 0\farcs8 window for the central region and a $3''$ window starting from a distance of $15''$ from
the center. The radial velocities were measured from the centroid positions of the Gaussians fitted in the emission lines. The accuracy of these measurements was estimated from the night-sky [OI]$\lambda6300$ line to be $\pm10\km$. We also measured the relative
intensities and FWHMs of the above emission lines.  The observed FWHMs were corrected for the instrumental profile width using the standard relation $(FWHM) ^2 =(FWHM) ^2_{obs}- (FWHM) ^2_{instr}$. 

To study the kinematics of the ionized gas and stars in the inner regions of UGC 5600 in detail, we observed the galaxy by the method of  integral-field spectroscopy with a multi-pupil fiber spectrograph (MPFS)  (Afanas 'ev et al. 2001)  attached to the 6m
telescope. The spectrograph simultaneously takes spectra from 240 spatial elements (constructed in
the form of square lenses) that form an array of $16\times 15$ elements in the plane of the sky. The angular
size of a single element is $1''$. A description of the spectrograph is given in the Internet on the SAO
www-page (\verb*"http://www.sao.ru/~gafan/devices.htm.").  Simultaneously with galaxy spectra, we took a 
night-sky spectrum from an area located at 4\farcm5 from the center of the field of view.
The detector was a TK1024 $1024\times1024$-pixel CCD array. The spectrograph reciprocal dispersion was $1.35$\AA\, per pixel and
the spectral resolution was $\sim3.5$\AA. The observations were performed sequentially in two spectral ranges. 
The "green" range included emission lines (\Hb, [O~III ]$\lambda\lambda4959, 5007$)  and absorption lines of the
galaxy stellar population (Mg~I$\lambda5175$ , Fe~I$\lambda5229$, Fe~I +Ca~I$\lambda5270$, etc.).  The "red "range contained the
\Ha , [N~II]$\lambda\lambda6548,6583$, [S~II]$\lambda\lambda 6716,6730$ emission lines.

\begin{figure}
\includegraphics[width=8  cm]{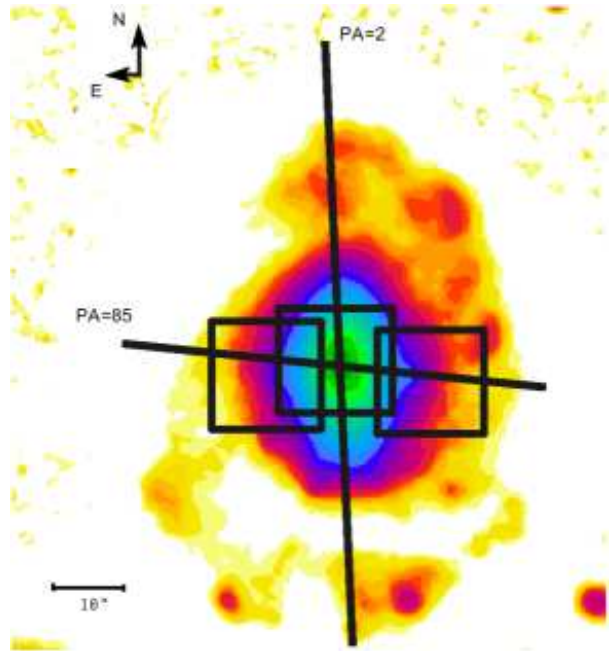}
 \protect\caption{
A B -band image of the galaxy UGC5600; the straight lines indicate the UAGS slit positions and the rectangles indicate
the MPFS fields. 
}
\end{figure}

\begin{table*}[t]
\caption{A log of observations of UGC 5600}
\begin{tabular}{r|c|l|l|l|l}
\hline
Date, instrument & Exposure time, s & Field & Seeing & Spectral range, \AA & PA, field \\
\hline
Jan. 28, 2000 &1800 &  $2''\times 140''$ & 2\farcs0 & $6200-7000$ & $85^\circ$ \\ 
UAGS             &1800  & $ 2 \times 140 $ & 2.0 & $ 6200-7000 $ & $0^\circ$\\ 
                       &1200 &  $2 \times 140 $ & 2.0 & $6200-7000 $ & $2^\circ$\\ 
Jan. 28, 2001 & 1800 & $16 \times 15 $ & 2.0 & $4300-5600$ & Center\\
MPFS             & 1200 & $16 \times 15 $ & 2.0 & $4300-5600$ & E side\\
                       &1200 & $16 \times 15 $ & 2.0 & $4300-5600$ & W side\\
                       &900 &  $16 \times 15 $ & 1.5 & $5550-6900$ & Center\\
                       &1200 & $16 \times 15 $ & 2.0 & $5550-6900$ & E side\\
                       &1200 & $16 \times 15 $ & 2.0 & $5550-6900$ & W side\\
Apr. 28, 2001  & 600 &  $16 \times15 $ & 2.0 & $4700-5900 $ &Center\\
MPFS               &1200 & $16 \times 15 $ & 2.0 & $4700-5900$ & $+5''$ to north\\
Aug. 11, 2001   &1200 & $16 \times 15 $ &1.0 & $4900-6200$ & Center\\
MPFS               &900 &   $16 \times 15 $ & 1.0 & $6000-7300$ & Center                \\
\hline
\end{tabular}
\end{table*}

We reduced the observations using the software developed at the SAO and running in the IDL 
environment. The primary reduction included bias subtraction, flat-fielding, cosmic-ray hit removal, extraction
of individual spectra from CCD images, and their wavelength calibration using the spectrum of a calibration lamp. 
Subsequently, we subtracted the nightsky spectrum from the galaxy spectra. The spectra 
of spectrophotometric standard stars were used to convert fluxes into absolute energies. 

We constructed two-dimensional intensity and radial-velocity (velocity fields)  maps in the \Ha, \Hb, 
[O~III], and [N~II] lines; the emission line profiles were also fitted with Gaussians. The
accuracy of the absolute radial-velocity determination estimated from sky lines ranges from 10 to $15\km$. 
The radial-velocity fields for the stellar component
were constructed by the cross correlation technique 
(Tonry and Davis 1979)  modified to work with integral-field 
spectroscopy and detailed by Moiseev (2001).  We
used the spectral range 5200 to 5500\AA\, containing
high-contrast lines of the galaxy stellar population. 
The spectra of G8-K3 III stars and the twilight sky
observed on the same nights as the galaxy were taken
as the radial-velocity standards. The accuracy of the
radial-velocity determination is $\sim10\km$. 

The January 2001 observations were carried out
at three different positions of the spectrograph field
of view (Fig.~1).  The resulting fields of velocities and
emission-line intensities were combined to give a
$40''\times16''$ total field of view. We measured the radial
velocities of the stellar component only for the central
$16''\times15''$ field, because the contrast of the stellar
lines in the outermost parts decreases sharply. 

In April 2001, we managed to construct a more extended radial-velocity field by using two MPFS fields
(one coincided with the galaxy photometric center and the other was displaced by $5''$ to the north).  The
resulting field of view was $16''\times20''$ . When studying
this velocity field in detail, we suspected that the central region of the galaxy, $r\sim2''$ in size, 
was kinematically decoupled (see below).  To check this feature, 
we repeated our observations of the stellar kinematics
in the central region of UGC 5600 in August 2001, 
at $\sim1''$ seeing. The derived $16''\times15''$ velocity field
with a higher angular resolution was also used in our
analysis. Here, all radial velocities were reduced to the
solar center (heliocentric velocities).

\section{RESULTS OF OBSERVATIONS
WITH THE LONG-SLIT SPECTROGRAPH}

Data on the UAGS observations are given in the
first three rows of the table. In our spectra of the
galaxy UGC~5600 along its major axis, the \Ha emission line is traceable up to distances 
of $\sim7-8$ kpc from the center.

\begin{figure}
\includegraphics[width=8  cm]{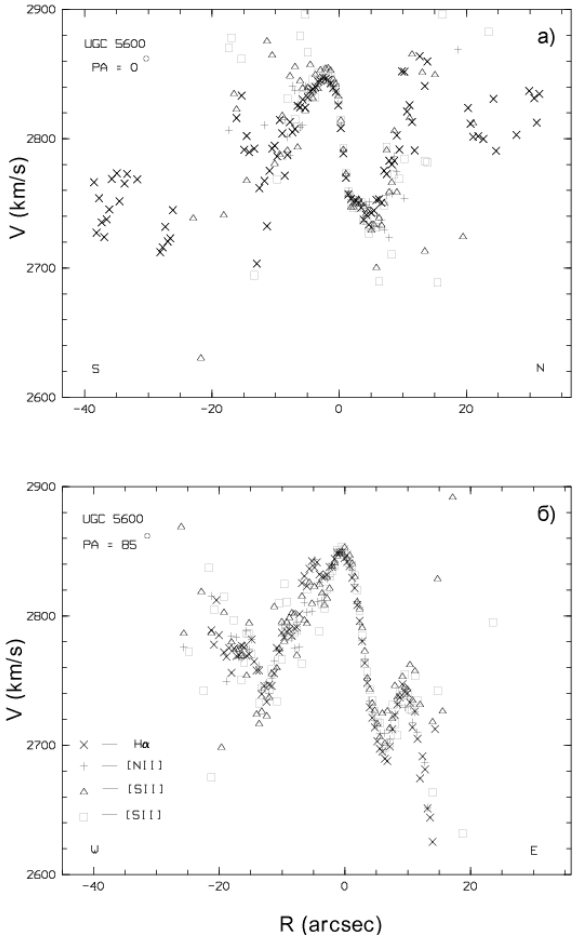}
 \protect\caption{
Radial-velocity curves (a) along the galaxy major axis and (b) along the ring major axis. 
}
\end{figure}

The radial-velocity curves along the galaxy major
axis ($P A =0$ )  and along the major axis of the inner
ring ($P A =85^\circ$ )  are shown in Fig. 2a and 2b. We see
from these figures that the radial velocities measured
from different  emission lines are equal, within the
error limits. Our radial-velocity curves are similar to
those in Reshetnikov and Combes (1994)  and the
small deviations are most likely due to difference  in
the spectrograph slit positions. The mean heliocentric
velocity of the photometric center is $2795\pm3\km$, 
which is lower than its value in Reshetnikov and
Combes (1994)  by $28\km$. 

The radial-velocity curve along the galaxy major
axis is complex in shape. The curve exhibits a small
rectilinear segment where the velocity increases from
0 to $55\km$; then, at a distance of $3''-4''$ from the
center, the scatter of points increases and further out, 
the mean radial velocity decreases. Reshetnikov and
Combes (1994) concluded that counterrotation was
observed in the central part of the galaxy ( $r \leq 5'' $).  
However, a detailed photometric study of this galaxy
(Karataeva et al., 2001) shows that we see the total
radiation from the galactic disk and inner ring exactly
at distances of $4''-10''$ from the center. Therefore, the
interpretation of the observed radial velocities in this
range is rather complex and contradictory.

\begin{figure*}
\includegraphics[width=15  cm]{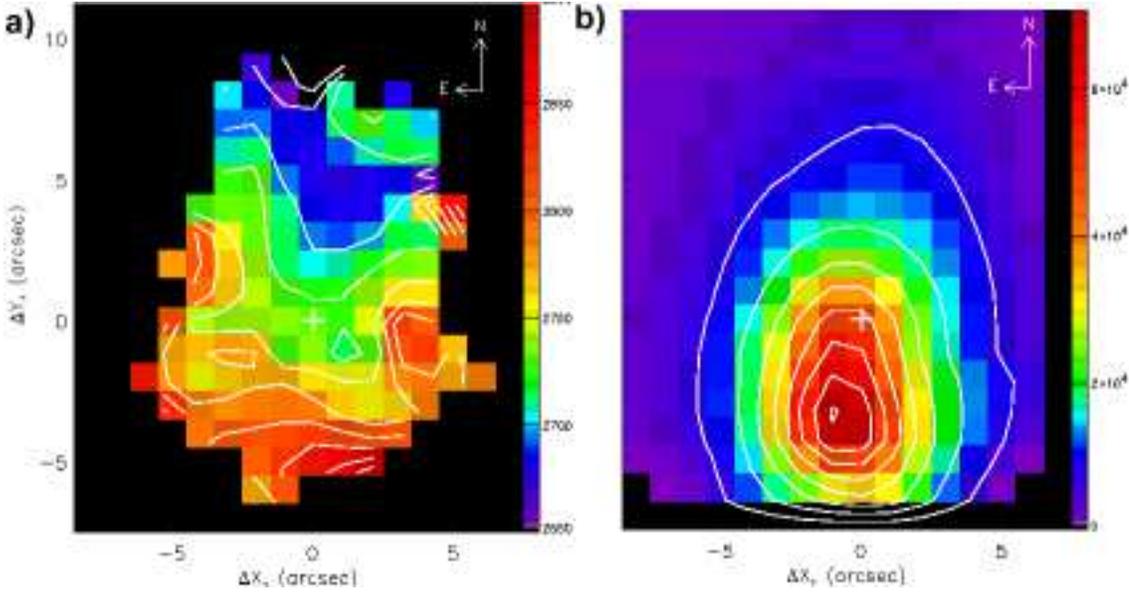}
 \protect\caption{
(a) The stellar radial-velocity field and (b) the continuum-flux distribution in the range $5200 -5450$\AA; the cross marks
the position of the kinematic center. 
}
\end{figure*}

Where the slit crosses the spiral arms $( r \geq  20'' )$,  
the measurement errors of the radial velocities are
large because the lines are weak. However, the velocity relative to the system center is, on average, 
$\km30\km$, with the southern side approaching us
and the northern side receding. 
The radial-velocity curve along the ring major axis (Fig. 2b) is also complex in shape. This is most
likely because the cut passes through different  galactic structures. 

Thus, we see that despite a wealth of information obtained with the long-slit spectrograph, the
interpretation of the observed radial velocities of the
emitting gas is ambiguous. One-dimensional cuts
are not enough to understand the kinematics of such a
multicomponent object. It is necessary to investigate
the two-dimensional velocity fields of the gas and
stars. This is the goal of our MPFS observations. 

\section{KINEMATICS OF THE STELLAR
AND GASEOUS COMPONENTS}

Data on the MPFS observations are also given
in the table. Below, we discuss the results of our 2D
spectroscopy. 

\subsection{The Radial-Velocity Distribution for the Stellar
Component}

We determined the radial velocities of the stellar
component by cross-correlation analysis. Since the
stellar-velocity dispersion turned out to be comparable
with the spectrograph instrumental profile $(\sigma\leq 70)$,  we could not reliably measure it and
study its variations across the galaxy. The low stellar velocity dispersion confirms the conclusion of 
Karataeva et al. (2001) that UGC5600 is a late-type galaxy. 

Figure 3 shows the stellar radial-velocity field and
the continuum ($5200 -5450$\AA) intensity distribution. 
We see from Fig. 3a that the isovels are complex
in shape. The isophotes in the continuum image
(Fig. 3b) are clearly distorted. These distortions may
be due to the presence of a feature in the nuclear
region, due to the clumpy structures in the ring
superimposed on the galaxy main body, and due to
the nonuniform distribution of dust, whose presence
follows from the IR fluxes (Richter et al. 1994),  both
in the ring and in the galactic disk. The presence of
dust may also affect the pattern of the radial-velocity
field. 

If we consider the central region belonging to the
galactic disk, then the shape of the isovels mainly 
corresponds to the circular rotation of stars around the
galaxy minor axis. For this region, we constructed the
average rotation curve and the radial dependence of
the kinematic-axis position angle. We used the tilted
ring technique (Begeman 1989; Moiseev and Mustsevoi 2000):
 the velocity field is broken down into
elliptical rings of fixed width and the rotation velocity
$V (r )$ and the kinematic-axis position angle $P A (r )$ 
are determined in each ring under the assumption of
circular rotation. In addition, conclusions about the
pattern of noncircular motions can be drawn from
an analysis of the variations in the position of the
kinematic axis and in the disk inclination to the line
of sight.

\begin{figure*}
\includegraphics[width=15  cm]{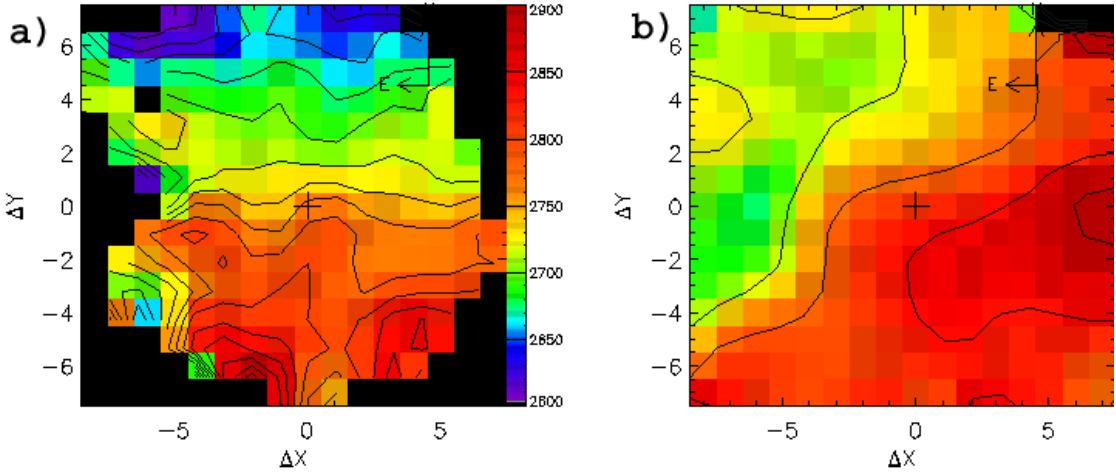}
 \protect\caption{
The radial-velocity fields for the galaxy central region: (a) for stars and (b) for gas, as constructed from \Ha at $1''$ seeing. 
}
\end{figure*}

The cross in Fig. 3 marks the position of the
kinematic center of the stellar component, which was
determined from the symmetry of the velocity field. 
The photometric and kinematic centers do not coincide,
 but the separation between them is less than
$2''$ . The differences in the positions of the photometric
and kinematic centers may be due to the presence
of features in the circumnuclear region. Within the
accuracy of our modeling, the position angle of the
kinematic axis $PA_{kin} =2^\circ$ coincides with the position angle of the photometric axis from Karataeva
et al. (2001) and the system heliocentric velocity
is $2740\pm15\km$ . The system velocities that we
determined from stars and gas differ approximately
by $50\km$ ; such differences are also observed
for other galaxies (see, e. g. , Sil 'chenko 1998).  The
inclination of the stellar disk to the plane of the sky
is $\sim50^\circ$, which is close to the value from Karataeva et
al. (2001),  and its rotation velocity slowly increases
from $40\km$ (at a distance of $2''$) to $90\km$ (at $8''$).  

Analyzing in detail the stellar radial-velocity
field in the circumnuclear region,  $\sim2''$ in size, we
suspected a peculiarity in the behavior of the isovels. 
However, the space resolution was too low to study it. 
Therefore, as we pointed out above, additional 
observations were carried out at seeing no worse than $1''$. 
Figure 4a shows the stellar radial-velocity field 
constructed from these data. Noncircular stellar motions
clearly show up in a $\sim2''$ region, which may suggest
the existence of a minibar that is possibly formed in
the disk because of the close passage of a companion
(Noguchi 1987).  The (nearly triangular) shape of the
isophotes in Fig. 3b and the asymmetry in the 
circumnuclear region of the photometric cuts along the
galaxy major axis shown in Fig. 2 from Karataeva et
al. (2001) are also consistent with this suggestion. 
However, observations with a higher space resolution
are needed for the final conclusion about the presence
of a minibar to be reached. 

Two features that recede with velocities
$\sim40 -60\km$ show up at $\sim$4\farcs5 from the center
in the NE and W directions (Fig. 3a).  The W feature
may belong to the inner ring. At the same time, the
velocity of the NE feature is opposite to the velocity of
the ionized gas in the inner ring, while this feature is
located far ($\sim2$ kpc) from the disk center. 

\subsection {The Radial-Velocity Distribution for the Gaseous
Component}

We constructed the radial-velocity fields of the gas
from hydrogen (\Ha , \Hb ) lines and from the forbidden
[O~III] line. They all proved to be similar. The
radial-velocity curves obtained from the velocity fields
are in good agreement with the radial-velocity curves
given in the preceding section. 

$\Ha$ is the brightest emission line. Since the 
accuracy of measuring the radial velocities from it is
higher, below we present only the data obtained from
this line. For convenience of comparing the behaviors
of the gaseous and stellar components, Fig. 4 shows
the velocity fields of the stars (a) and the ionized gas
(b) for the galaxy central region. A detailed 
analysis of the gas kinematics indicates that the slope
of the isovels in the central part,  $\sim3$ in diameter, 
is identical to the slope of the isovels for the stellar
component. Further out, in the E -W direction, the
isovels are turned through about $90^\circ$, suggesting gas
rotation about the galaxy major axis. 

To analyze the behavior of the emitting gas in the
galactic inner ring, Fig. 5 shows the \Ha intensity
distribution, the total radial-velocity field, and 
continuum ($6100 -6300$ \AA) intensity variations. In contrast
to the continuum intensity distribution (Fig. 5c),  the
\Ha image (Fig. 5a) is elongated from east to west and
coincides with the ring location, but the ellipticity of
the \Ha isophotes changes. They become rounder as
one approaches the center. South of the center, the
isophotes flatten. This isophotal behavior can be explained 
by assuming that there are two gaseous components. The first is associated with the galactic disk
and the second is associated with the inner ring. In
the central region, we observe the total radiation from
the two components. The isophote flattening south
of the center may imply that on this side the ring is
projected onto the galactic disk and its dust partially
absorbs the disk radiation. Note that our assumption
is in conflict with the conclusion of Karataeva et al. 
(2001).  Since the region with blue color indices on the
southern side is narrower than on the northern side, 
these authors assumed that the ring was projected
onto the galactic body north of the center and was
seen through it south of the center. However, such
a peculiarity of the color indices may be due to the
nonuniform ring structure. For example, it may stem
from the fact that the southern side of the ring is
slightly narrower than its northern side. Therefore, the
behavior of the \Ha isophotes seems to characterize
the ring orientation more reliably, especially since the
isophotes are similar in shape in all emission lines. 

The existence of two gaseous components must
affect the shape of the line profiles along the galaxy
major axis, particularly at distances of $5'' -10''$ south
and north of the center, where the ring and the disk
are superimposed on each other. A significant scatter
of points is observed precisely in these segments of
the radial-velocity curve (Fig. 2).  The line profiles are
irregular in shape and can be fitted by two Gaussians. 
However, the reliability of this fit is low, because the
errors are large. 

The assumption that there are two gaseous components is also confirmed 
by the shape of the isovels
in Figs. 4b and 5b. At the center, where the directions
of the isovels for the gaseous and stellar components
coincide, the radial-velocity field is determined by the
motion of the gas belonging to the galactic disk, while
starting approximately from $2''$ east and west of the
center, the ring gas motion shows up clearly.

\begin{figure*}
\includegraphics[width=15  cm]{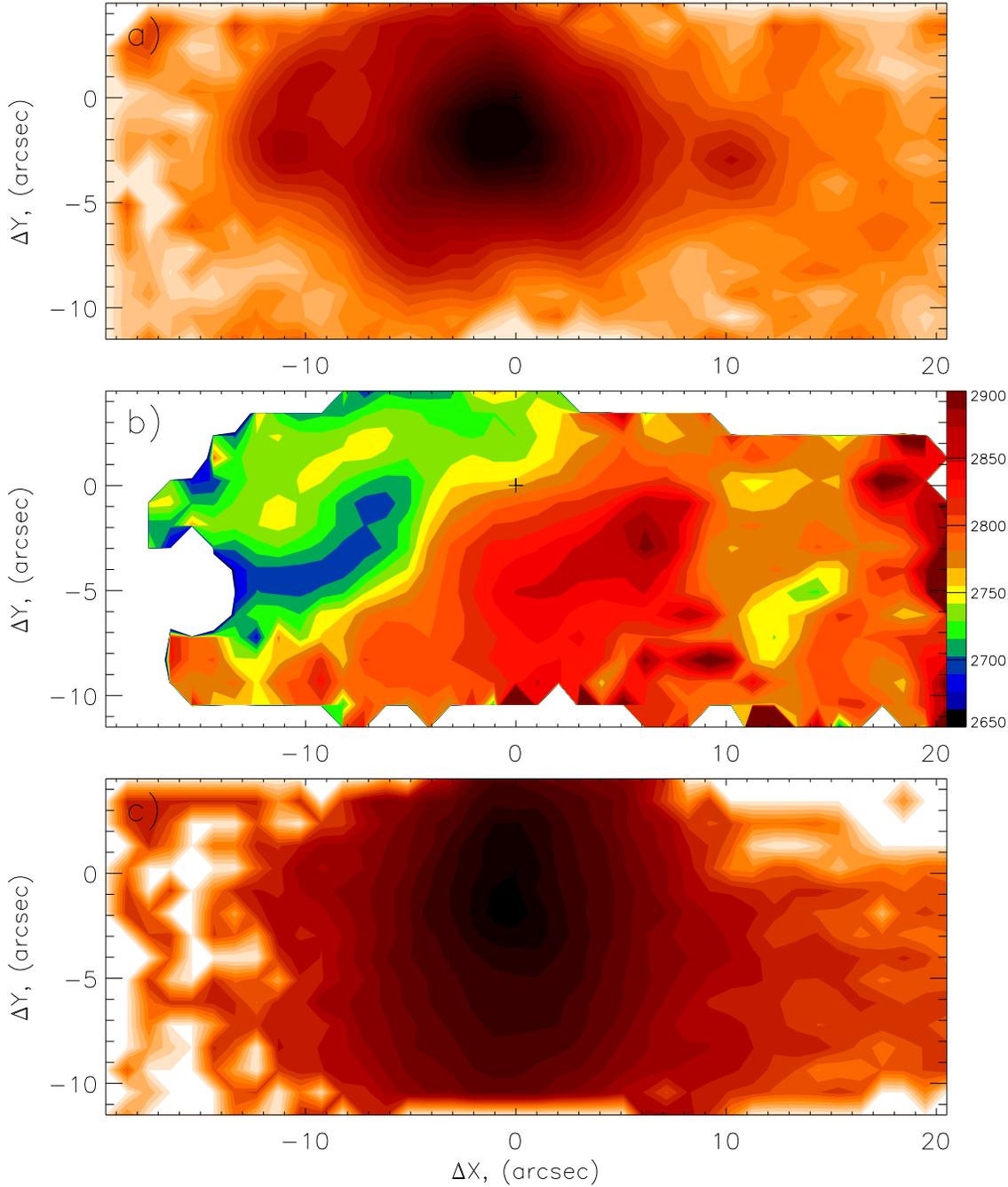}
 \protect\caption{
(a) The intensity distribution, (b) the radial-velocity field in \Ha , and (c) the continuum intensity distribution in the
range $6100 -6300$\AA (the intensities are on a logarithmic scale).  
}
\end{figure*}

Assuming circular rotation, we attempted to
model the gas motion in the ring (without the central
region $\sim3''$ in size) and to estimate the ring inclination 
to the plane of the sky. We estimated the latter to
be $\sim70^\circ-75^\circ$. Knowing the inclination of the galactic
disk and the ring to the plane of the sky, we can
determine the angle between the disk and the ring
from the relation
$$
\cos\Delta i = \pm\sin i_1 \sin i_2 \cos (P A_1-PA_2) +\cos i_1\cos
i_2~~(1)
$$
where $i_1$ and $i_2$ are the disk and ring inclinations to
the plane of the sky, $P A_1$ and $P A_2$ are the position
angles of the major axes of the galactic disk and the
ring. This angle was found to be about $78^\circ\pm5^\circ$; i. e. , 
the ring is polar. 
A close examination of Fig. 5b reveals several
features in the radial-velocity field. For example, there
is a feature receding with a velocity of $\sim100\km$ at
distance of $7''$ to the west. Such features may result
from the nonuniform ring structure. Their presence
may stem from the fact that individual bright H\,II
regions fall on the line of sight. 
The right part of Fig. 5b, starting from $17'' -20''$, 
corresponds to the spiral arm receding at a velocity of
$\sim30 -40\km$.

\begin{figure}
\includegraphics[width=8  cm]{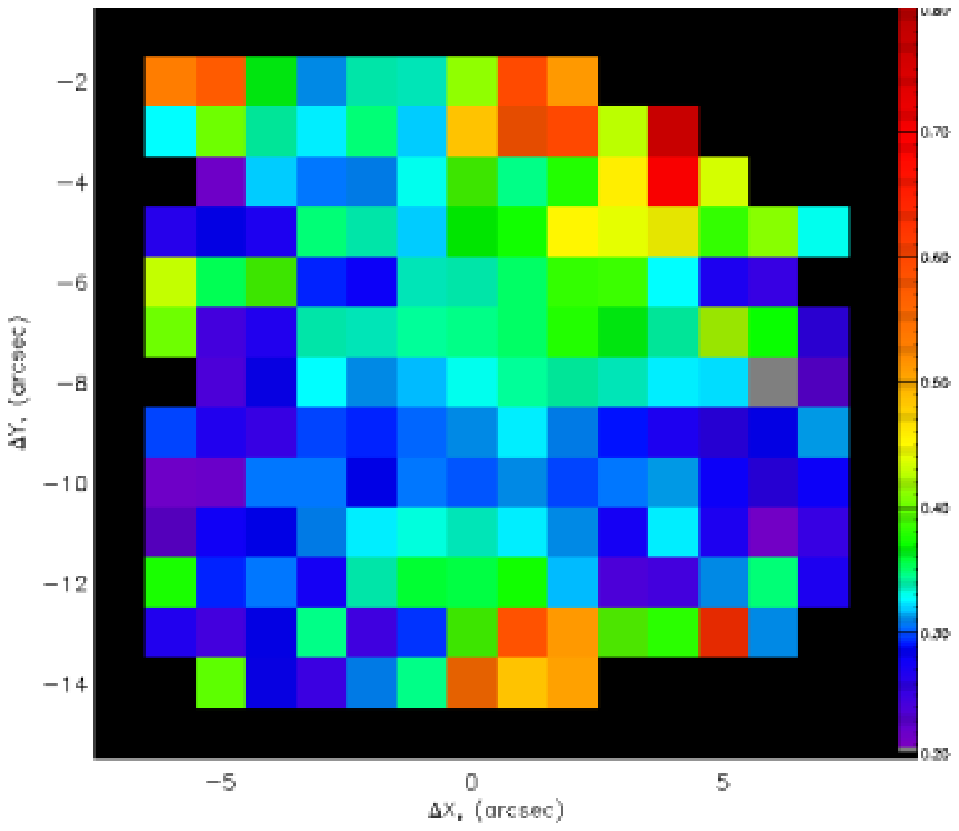}
 \protect\caption{
The distribution of the $\mbox{[N~II]}/$\Ha intensity ratio in the galaxy $14\times13''$ central region. 
}
\end{figure}

\section{CHARACTERISTICS
OF THE EMISSION-LINE REGIONS}

As was already pointed out above, we measured
the \Ha FWHMs and the relative emission-line 
intensities at various distances from the nucleus. 
Our measurements for the nuclear region closely agree, within
the error limits, with the results of Reshetnikov and
Combes (1994),  who showed that the nuclear 
emission originates from H II regions. Figure 6 presents
the distribution of the $\mbox{[N~II]}/$\Ha ratio for the central
region of the galaxy. The logarithms of this ratio fall
within the range $-0.6$ to $-0.35$ , which is characteristic 
of H~II regions (Veilleux and Osterbrock 1987).  
Therefore, we may conclude that the emission lines in
the nucleus and in the ring originate in H~II regions. 
The intensity ratio of forbidden and permitted lines is
virtually constant along the major axis of the polar
ring, suggesting that the physical conditions in the
emission-line regions are similar. The forbidden-line
intensity increases with distance from the nucleus
along the galaxy major axis compared to \Ha . To
confirm that the increase in forbidden-line intensity
(Fig. 6) is unrelated to the edge effects of the multipupil 
spectrograph, we considered the data acquired
with the long-slit spectrograph. It turned out that the
$\mbox{[N~II]}/$\Ha ratio at the center is 0.35 and then gradually 
increases, reaching 0.6 at $r\sim5''-7''$; further out, it
is roughly constant up to $10''-12''$. The emission-line
intensity decreases with distance from the center and
starting from about $14''$, the signal-to-noise ratio is
$\leq 3$ . Therefore, we cannot reliably determine the line
intensity ratio in these regions. 

\begin{figure}
\includegraphics[width=8  cm]{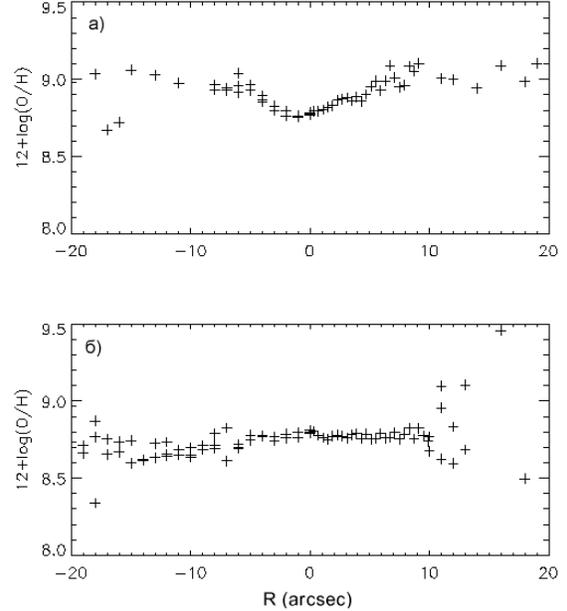}
 \protect\caption{
Variations in the oxygen abundance (a) along the
galaxy major axis and (b) along the major axis of the polar ring. 
}
\end{figure}

The strengthening of the nitrogen forbidden lines
compared to \Ha appears to be due to an increase
in the importance of the collisional excitation as the
galactic gaseous disk interacts with the polar-ring
gas. 

Here, our prime objective was to study the 
kinematics of the gaseous and stellar components. Therefore, 
the observations were carried out only in the
green and red spectral ranges and we cannot determine 
the physical conditions in H II regions and their
chemical composition from our data. However, we
attempted to estimate the metallicity in the emission-line regions. 

According to Denicolo et al. (2001),  there is an
empirical relation between $\log (\mbox{[N~II]}\lambda6583/\mbox{H}_\alpha)$ and
$\log(\mbox{O}/\mbox{H})$: 
$$
\begin{array}{c}
12 +\log(\mbox{O}/\mbox{H}) =9.12(\pm0.05) +0.73(\pm0.10) \\ 
\,\:\times\log (\mbox{[N~II]}\lambda6583/\mbox{H}_\alpha)~~~~~~~(2); 
\end{array}
$$
its form depends neither on the flux calibration nor
on the reddening corrections. Using this relation, we
estimated the metallicity in H~II regions. Figure 7
shows the radial distribution of $12 +\log(\mbox{O}/\mbox{H})$ along
the galaxy major axis and along the ring major axis. 
As was already pointed out above, the shape of the
distribution in Fig. 7a results from an increase in the
nitrogen-line intensity compared to the \Ha intensity, 
possibly because the collisional excitation increases
in importance. Besides, the empirical relation  is
reliable for an $\mbox{[N~II]}\lambda6583/\mbox{H}_\alpha)$ ratio below 0.5.
 Therefore, the increase in $12 +\log(\mbox{O}/\mbox{H})$ in the regions between
$5''$ and $10''$ north and south of the center may not
result from an increase in metallicity. The values of
$12 +\log(\mbox{O}/\mbox{H})$ are virtually constant along the major
axis of the polar ring (Fig. 7b),  suggesting that the
physical conditions in the ring are homogeneous. 

The mean metallicity in the circumnuclear region
and in the polar ring is $\sim8.8$ , which corresponds to
$0.9Z_\odot$ . A similar estimate was also obtained from
[S~II] lines by using the sum of the [S~II]$\lambda6717 + \lambda6731$ line fluxes.
 This result corresponds to normal
evolution in galaxies with such luminosities (Richer
et al. 1998).  

The nearly solar metallicity in the polar ring implies 
that it cannot not be produced by the capture of
a dwarf companion. 

\section{CONCLUSIONS}

In conclusion, we summarize our main results. 
\begin{enumerate}

\item Based on 2D spectroscopy, we constructed
the radial velocity fields of the stellar and gaseous
components for the central regions of the peculiar
galaxy UGC 5600. 

\item An analysis of these fields revealed two kinematic 
subsystems: the first is related to the galactic
disk and the second is related to the inner ring. 

\item The angle between the disk and ring planes
was found about $78^\circ\pm5^\circ$; i.e., the inner
ring is polar. This provides compelling evidence that
the galaxy UGC 5600 belongs to PRGs. 
\item We established from the intensity ratio of forbidden 
and permitted lines that the emission originates 
in H II regions; the metallicity was estimated. 
It proved to be nearly solar, which rules out a dwarf
galaxy as the donor in forming the polar ring. 
\end{enumerate}

\begin{acknowledgements}
We are grateful to the 6m telescope committee
for allocating telescope observational time and to the
SAO staff: V. L. Afanasiev for assistance in the MFPS
observations and for providing data reduction software; 
A. N. Burenkov for assistance in the UAGS
observations and data reductions. We wish to thank
O. K. Sil 'chenko for helpful discussions and remarks. 
This study was supported in part by the "Astronomy "
program (project no.1.2.6.1) and the "Integration "
program (project no.A0007).  
\end{acknowledgements}

\bigskip
{\it Translated by N. Samus'}
\end{document}